\documentclass[aps,pra,floatfix,11pt]{revtex4-2}
\usepackage{graphicx} % Required for inserting images
\usepackage{comment}
\usepackage{xcolor}
\usepackage{braket}
\usepackage{graphics,graphicx,epsfig}
\usepackage[caption=false]{subfig}
\usepackage{amssymb,amsmath}
\usepackage{epstopdf}
\usepackage{amsmath}
\usepackage{mathtools}
\def\d{{\rm d}}
\begin{document}
\title{Quadratic and cubic scrambling in the estimation of two successive phase-shifts}
\author{Manju}
\email{manju@unimi.it}
\author{Stefano Olivares}%
\email{stefano.olivares@fisica.unimi.it}
\affiliation{% 
Dipartimento di Fisica ``Aldo Pontremoli'', Universit\`a degli Studi di Milano, I-20133 Milano, Italy
}%
\author{Matteo G. A. Paris}%
\email{matteo.paris@fisica.unimi.it}
\affiliation{% 
Dipartimento di Fisica ``Aldo Pontremoli'', Universit\`a  degli Studi di Milano, I-20133 Milano, Italy
}%
\date{\today}
\begin{abstract}
Multiparameter quantum estimation becomes challenging when the parameters are incompatible, i.e., when their respective symmetric logarithmic derivatives do not commute, or when the model is sloppy, meaning that the quantum probe depends only on combinations of parameters leading to a degenerate or ill-conditioned Fisher information matrix. 

In this work, we explore the use of scrambling operations between parameter encoding to overcome sloppiness. We consider a bosonic model with two phase-shift parameters and analyze the performance of second- and third-order nonlinear scrambling using two classes of probe states: squeezed vacuum states and coherent states. Our results demonstrate that nonlinear scrambling mitigates sloppiness, increases compatibility, and improves overall estimation precision. We find third-order nonlinearity to be more effective than second-order under both fixed-probe and fixed-energy constraints. Furthermore, by comparing joint estimation to a stepwise estimation strategy, we show that a threshold for nonlinear coupling exists. For coherent probes, joint estimation outperforms the stepwise strategy if the nonlinearity is sufficiently large, while for squeezed probes, this advantage is observed specifically with third-order nonlinearity.
\end{abstract}
\maketitle
\section{INTRODUCTION}
Quantum metrology leverages quantum resources to surpass the precision limits of classical sensing \cite{paris2009quantum,giovannetti2006quantum, giovannetti2011advances,toth2014quantum,albarelli2020perspective}. While the single-parameter case is well-understood, with its ultimate precision governed by the quantum Cram\'er-Rao bound (QCRB), the simultaneous estimation of multiple parameters presents a far richer set of problems and challenges \cite{szczykulska2016multi, albarelli2020perspective}. As a matter of fact, multiparameter estimation is not merely a technical extension and in many applications, from characterizing complex biological samples \cite{transtrum2015perspective} to imaging and gravitational-wave detection \cite{schnabel2010quantum, danilishin2012quantum}, measuring several quantities at once is essential and can, in principle, yield a fundamental advantage over estimating them individually.

However, this potential is hampered by two distinct challenges. First, at the measurement stage, 
the best measurement strategy for one parameter may collide with the best strategy for another. This \emph{incompatibility} arises when the optimal measurements for different parameters do not commute \cite{ragy2016compatibility, genoni2013optimal, liu2020quantum}. This quantum mechanical trade-off 
prevents the simultaneous saturation of the QCRB for all parameters.  Second, there 
is a challenge at the encoding stage: \emph{sloppiness} \cite{brown2003statistical, waterfall2006sloppy, machta2013parameter,goldberg2021taming,yang2023untwining,he2025scrambling}. A model is sloppy when different combinations of parameters produce nearly identical output states, making them practically indistinguishable. This results in a singular or ill-conditioned quantum Fisher information matrix (QFIM), indicating that the probe state is not sensitive to changes in individual parameters, but only to certain combinations of them. Sloppiness thus acts as a source of noise, fundamentally limiting estimation precision.

While incompatibility has been widely studied \cite{zhu2015information,heinosaari2016invitation,ragy2016compatibility,chen2022incompatibility,xia2023toward,candeloro2024dimension}, the problem of sloppiness, particularly in continuous-variable (CV) systems relevant to quantum optics, has received less attention. 
Motivated by recent advances in this context \cite{sharma2025mitigating, frigerio2025overcoming,he2025scrambling}, here we investigate a canonical example of a sloppy model involving 
the sequential application of two phase shifts. In this case, the probe state is only sensitive to the total phase, rendering the individual parameters unidentifiable. We demonstrate how to actively \emph{engineer} the quantum statistical model to overcome its intrinsic sloppiness. Upon introducing a nonlinear operation between the two successive phase shifts, the information about the individual parameters is \emph{scrambled} throughout the Hilbert space. We analyze and compare the efficacy of second-order (quadratic) and third-order (cubic) nonlinearities in mitigating sloppiness for both coherent and squeezed vacuum probe states.  
Our results show that nonlinear scrambling helps to separate the effects of each parameter, mitigating sloppiness, increasing compatibility, and overall improving estimation accuracy. Third-order nonlinearity is more effective at fixed probe and energy. We also compare joint estimation to a step-wise one, discussing in which regimes joint estimation is outperforming step-wise estimation.

The paper is structured as follows: In Section~\ref{s:MPQM}, we provide a brief overview of the key concepts and theoretical tools of the multi-parameter quantum metrology that are required for our problem. In Section~\ref{s:scrambling}, we introduce our estimation scheme and the scrambling technique to reduce the sloppiness in the model. We also optimize the relevant bounds to precision and examine the role of sloppiness in achieving these bounds and compare joint estimation strategies with step-wise ones. Section~\ref{s:concl} closes the paper with some concluding remarks.

\section{MULTI-PARAMETER QUANTUM METROLOGY}\label{s:MPQM}

In this section, we will provide all the basic notions of multi-parameter quantum metrology that are needed for our objectives (for further details see Refs.~\cite{albarelli2020perspective,szczykulska2016multi,razavian2020quantumness}). In a multi-parameter quantum metrological problem, the goal is to jointly estimate a set of parameters \(\phi = \{\phi_1, \ldots, \phi_N\}\) where \(N > 1\) encoded in a quantum state \(\rho_{{\phi}}\), 
%This family of states, \(\rho_{{\phi}}\), is
typically referred to as a quantum statistical model.
To infer the values \(\{\phi_\mu\}_{\mu}\), we perform a quantum measurement described by a positive operator-valued measure (POVM) \(\{\Pi_x\}_x\), which satisfies \(\Pi_x \geq 0\) and \(\int \d x \, \Pi_x = \mathbb{\hat{I}}\), where \(\mathbb{\hat{I}}\) is the identity operator over the entire Hilbert space. By repeating the measurement \(M\) times, we obtain a statistical sample of independent and identically distributed outcomes \({x} = \{x_1, \ldots, x_M\}\), from which parameter estimates are obtained through an estimator function \(\hat{{\phi}} = \hat{{\phi}}({x})\). Given this, the task is to determine the optimal POVM that achieves the highest accuracy in the estimation of \({\phi}\), i.e., the one minimizing the uncertainty as much as possible.

For unbiased estimators, i.e., its expected value is equal to the true value of the parameter, namely, \(E[\hat{{\phi}}] = {\phi}\), the accuracy is quantified by the covariance matrix \(V\) associated with \(\hat{{\phi}}\):
\begin{equation}
V(\hat{{\phi}}) = \int \d x \, p(x|{\phi}) \left[ \hat{{\phi}}(x) - {\phi} \right] \left[ \hat{{\phi}}(x) - {\phi} \right]^T,
\end{equation}
where \(p(x|{\phi}) = \prod_{j=1}^M p(x_j|{\phi})\) and \(p(x|{\phi}) = \mathrm{Tr}[\rho_{{\phi}} \Pi_x]\) is the conditional probability of obtaining outcome \(x\) given \({\phi}\). The covariance matrix satisfies the classical Cramér-Rao (CR) bound, defined by the following matrix inequality:
\begin{equation}
V(\hat{{\phi}}) \geq \frac{1}{M} \mathcal{F}^{-1}({\phi}),
\label{1}
\end{equation}
where the \(\mathcal{F}({\phi})\), known as the classical Fisher information matrix (FIM), associates a positive-definite invertible matrix with
each probability distribution in the statistical model whose elements \(\mathcal{F}_{\mu\nu}({\phi})\) are as follows:
\begin{equation}
\mathcal{F}_{\mu\nu}({\phi}) = \int \d x \, p(x|{\phi}) \left[ \partial_\mu \log p(x|{\phi}) \right] \left[ \partial_\nu \log p(x|{\phi}) \right],
\end{equation}
with \(\mu, \nu = 1, \ldots, N\), and \(\partial_\mu\) denotes partial derivatives with respect to \(\phi_\mu\). Moreover, an asymptotically efficient estimator, which reaches the bound \((\ref{1})\) for large sample sizes \(M \rightarrow +\infty\), is always attainable.Both the maximum-likelihood estimator and the Bayesian estimator offer this asymptotic efficiency \cite{albarelli2020perspective}. However, the FIM depends on the specific POVM \(\{\Pi_x\}_x\) being implemented, thus it is usually considered as a classical quantity. As a consequence, a more general bound can be obtained by optimizing the FIM over all possible quantum measurements, leading to a quantum version of the CR bound that only depends on the considered statistical model \(\rho_{{\phi}}\).

In the single-parameter scenario, this problem was exactly solved by Helstrom, who introduced the quantum Fisher information as the relevant figure of merit, obtained as the maximum Fisher information over all POVMs. In the multi-parameter setting, there exist different possible approaches, corresponding to different figures of merit. If we employ the symmetric logarithmic derivatives (SLD) operators \(L_\mu\), \(\mu = 1, \ldots, N\), defined via the Ljapunov equation \cite{helstrom1969quantum}:
\begin{equation}
\partial_\mu \rho_{{\phi}} = \frac{1}{2} \left( L_\mu \rho_{{\phi}} + \rho_{{\phi}} L_\mu \right),
\end{equation}
the QFIM \(Q({\phi})\) has elements:
\begin{equation}
Q_{\mu\nu} = \mathrm{Tr} \left[ \rho_{{\phi}} \frac{L_\mu L_\nu + L_\nu L_\mu}{2} \right],
\end{equation}
%We re-express Equation (5) in a more manageable way by considering the spectral decomposition of \(\rho_{{\phi}}\), \(\rho_{{\phi}} = \sum_k \rho_k |\lambda_k\rangle \langle \lambda_k|\), that, combined with (4), leads to:
%\begin{equation}
%Q_{\mu\nu} = 2 \sum_{k,j} \frac{\langle \lambda_k | \partial_\mu \rho_{{\phi}} | \lambda_j \rangle \langle \lambda_j | \partial_\nu \rho_{{\phi}} | \lambda_k \rangle}{\rho_j + \rho_k}.
%\end{equation}

The QFIM provides a tighter matrix lower bound on Eq.~(\ref{1}), referred to as the SLD-quantum Cramér-Rao (SLD-QCR) bound:
\begin{equation}
V(\hat{{\phi}}) \geq \frac{1}{M} Q^{-1}({\phi}).
\label{CQ}
\end{equation}
These matrix inequalities can be turned into scalar bounds by introducing a semipositive definite \(N \times N\) weight matrix \(W\); then we have \cite{albarelli2020perspective}: \(\mathrm{Tr}[W V] \geq C_\mathcal{F}(W)\) and \(\mathrm{Tr}[W V] \geq C_Q(W)\), with \(C_\mathcal{F}(W) = M^{-1} \mathrm{Tr}[W \mathcal{F}^{-1}]\) and \(C_Q(W) = M^{-1} \mathrm{Tr}[W Q^{-1}]\).

However, unlike the single-parameter scenario where the QCR bound may be achieved by a projective measurement over the SLD eigenstates, in the multi-parameter setting, the SLD-QCR bound (\ref{CQ}) is not attainable in general, as the SLDs associated with the different parameters may not commute with one another. In this case, the parameters are incompatible, and there is no joint measurement that allows one to estimate all the parameters with the ultimate precision. Accordingly, one may introduce two other relevant bounds.

The first bound, referred to as the most informative bound, corresponds to \(C_{\text{MI}}(W) = M^{-1} \min_{\text{POVM}} \{\mathrm{Tr}[W \mathcal{F}^{-1}]\}\), which, in general, does not coincide with the SLD-QCR bound in the presence of multiple parameters. The second one is the so-called Holevo Cramér-Rao (HCR) bound \cite{holevo1977commutation} \(C_{\text{H}}(W)\), which turns out to be the most informative bound of the asymptotic statistical model, i.e., the minimum FI bound achieved by a collective POVM performed on infinitely many copies of the statistical model \cite{albarelli2020perspective,razavian2020quantumness}, namely \(\rho_{{\phi}}^{\otimes n}\) with \(n \gg 1\). Thereafter, we have
\begin{equation}
  \mathrm{Tr}[W V] \geq C_\mathcal{F}(W) \geq C_{\text{MI}}(W) \geq C_{\text{H}}(W) \geq C_Q(W), 
\end{equation}
and, thus, the HCR bound is usually regarded as the most fundamental scalar bound for multi-parameter quantum estimation. Given this hierarchy, the compatibility of parameters is achieved, at least asymptotically, when the HCR bound saturates the SLD-QCR limit. To this aim, it has been recently proved that:
\begin{equation}
C_Q(W) \leq C_{\text{H}}(W) \leq (1 + R) C_Q(W),
\label{CQR}
\end{equation}
where the quantumness parameter \(R\) is given by \cite{carollo2020geometry, carollo2019quantumness, candeloro2021properties}:
\begin{equation}
R = \| i Q^{-1} D \|_\infty,
\end{equation}
in which \(\| A \|_\infty\) denotes the largest eigenvalue of the matrix \(A\), and \(D\) is the asymptotic incompatibility matrix, also referred to as Uhlmann curvature \cite{carollo2020geometry}, with matrix elements:
\begin{equation}
D_{\mu\nu} = -\frac{i}{2} \mathrm{Tr} \left[ \rho_{{\phi}} (L_\mu L_\nu - L_\nu L_\mu) \right],
\end{equation}
The quantumness parameter \(R\) satisfies \(0 \leq R \leq 1\) and \(R = 0\) iff \(D({\phi}) = 0\); therefore, it provides a measure of asymptotic incompatibility between the parameters. An important feature of the quantumness parameter \(R\) is its independence from the weight matrix \(W\), ensuring that bound in Eq. (\ref{CQR}) remains valid for any choice of \(W\). However in many practical scenarios, a natural weight matrix arises from the intrinsic characteristics of the quantum statistical model. In such cases, one can derive a weight-dependent hierarchy of inequalities for the Holevo bound \(C_{H}\), which is tighter than that in Eq.~(\ref{CQR}) \cite{carollo2019quantumness,albarelli2019upper,tsang2019holevo}. We have
\begin{align}
    C_Q( W) &\leq C_{H}(W) \leq (1+T_{W})C_Q(W) \equiv C_T(W),
    \label{CH1}
\end{align}
where 
\begin{equation}
T_{W}=\frac{\parallel \sqrt{W} Q^{-1} D Q^{-1} \sqrt{W} \parallel_{1}}{C_{Q}},
\end{equation}
%D_{\mu\nu} = 4 \sum_{k,j} \frac{\rho_k}{(\rho_k + \rho_j)^2} \mathrm{Im} \left[ \langle \lambda_k | \partial_\mu \rho_{{\phi}} | \lambda_j \rangle \langle \lambda_j | \partial_\nu \rho_{{\phi}} | \lambda_k \rangle \right].
%\end{equation}
and \(\| A \|_1\) denotes the sum of the singular values of \(A\). Equations (\ref{CQR}) and (\ref{CH1}) implies that the SLD-QCR bound is saturated iff \(D\) is the null matrix, referred to as the \emph{weak compatibility condition}, and the parameters are said to be asymptotically compatible. 
%In addition, the quantity \(R\) satisfies \(0 \leq R \leq 1\) and \(R = 0\) iff \(D({\phi}) = 0\); therefore, it provides a measure of asymptotic incompatibility between the parameters. 
In particular, for \(N = 2\) parameters, \(R\) reduces to:
\begin{equation}
R = \sqrt{\frac{\det D}{\det Q}} = \sqrt{\frac{S}{C}}
\end{equation}
where \(S\) quantifies the sloppiness of the model, and \(C\) measures the compatibility between the parameters. 
\begin{equation}
C = \frac{1}{\det D}, \quad
S = \frac{1}{\det Q},
\label{sc}
\end{equation}
%On the other hand, in multiparameter quantum estimation, the QCRB cannot, in general, be saturated due to the non-commutativity of the SLDs associated with different parameters. Based on the framework of quantum local asymptotic normality \cite{hayashi2008asymptotic,yamagata2013quantum,kahn2009local}, it has been shown that the SLD-based multiparameter QCRB is attainable if and only if the weak compatibility condition is satisfied \cite{ragy2016compatibility}, given by:
%\begin{equation}
%\mathrm{Tr}\left( \rho_\phi [L^S_\mu, L^S_\nu] \right) = 0, 
%\end{equation}
%where \( L^S_\mu \) and \( L^S_\nu \) are the SLDs corresponding to the parameters \( \phi_\mu \) and \( \phi_\nu \), respectively.
In case of two-parameter pure probe state models, analytical expressions of the bounds becomes available since the QFI matrix elements take an explicit form given by
\begin{equation}
\label{A}
Q_{\mu\nu} = 4 \, \mathrm{Re} \left[ \langle \partial_\mu \psi_{{\phi}} | \partial_\nu \psi_{{\phi}} \rangle - \langle \partial_\mu \psi_{{\phi}} | \psi_{{\phi}} \rangle \langle \psi_{{\phi}} | \partial_\nu \psi_{{\phi}} \rangle \right].
\end{equation}
%To quantify the degree of incompatibility between pairs of parameters, one introduces the incompatibility matrix \( D \), also known as the Uhlmann curvature, defined by equation (12). For a two-parameter pure state model,
and the analogous expression for the elements of the Uhlmann curvature is given by
\begin{equation}
   \quad D_{\mu\nu}  = 4\, \mathrm{Im} \left[ \langle \partial_\mu \psi_\phi | \partial_\nu \psi_\phi \rangle - \langle \partial_\mu \psi_\phi | \psi_\phi \rangle \langle \psi_\phi | \partial_\nu \psi_\phi \rangle \right].
   \label{B}
\end{equation}

\section{SCRAMBLING, PRECISION, SLOPPINESS AND INCOMPATIBILITY}\label{s:scrambling}
\begin{figure}[h!]
	\centering
	\includegraphics[width=.8\textwidth]{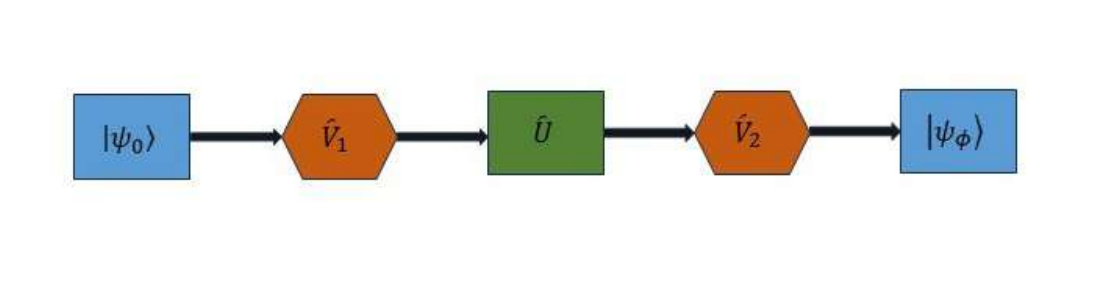}
	\caption{The sloppy model considered in this paper involves encoding the model parameters $\phi_1$ and $\phi_2$ via the unitary operations $\hat{V}_1$ and $\hat{V}_2$. To remove sloppiness, we introduce a scrambling operation, represented by the intermediate unitary transformation $\hat{U}$.}
	\label{Model}
\end{figure}

To systematically investigate the sloppiness and incompatibility in a CV quantum statistical model, we consider the two-parameter CV model schematically illustrated in Fig. \ref{Model}.
%By introducing a tunable scrambling operation during parameter encoding, we control the correlations between parameters and quantify their impact on estimation precision. 

Owing the convexity of the QFIM, we restrict our analysis to a pure probe state \( |\psi_0\rangle \). The parameters to be estimated are the two phases \(\phi_{1}\) and \(\phi_{2}\), which are encoded onto the input state via unitary transformations
\begin{equation}
  \hat{V}_{k} = e^{-i \phi_{k} \hat{n}},  \quad k=1,2  
\end{equation}
where \(\hat{n}= \hat{a}^{\dagger}\hat{a}\) is the number operator and \( \hat{a} \) and \( \hat{a}^\dagger \) are the annihilation and creation operators, respectively. We considered a generic unitary scrambler \(\hat{U}\), which rotates the basis states between the two phase encoders. When the unitary scrambler in between the two phases is absent, the model becomes sloppy the two parameters contributing to an overall phase shift of \(\phi_{1}+\phi_{2}\) of the input state. If we insert a scrambling operation  \(\hat{U}\) the output state is given by 
\begin{equation}
    |\psi\rangle =  \hat{V}(\phi_{2}) \hat{U}(\gamma) \hat{V}(\phi_{1}) |\psi_{0}\rangle
\end{equation}
and the entries of QFIM may be evaluated using Eq. \((\ref{A})\), leading to
\begin{align}
& Q_{11} = 4 \, \mathrm{Re}\left[\langle \psi_{0}|\hat{n}^{2}|\psi_{0}\rangle - \big(\langle \psi_{0}|\hat{n}|\psi_{0}\rangle\big)^{2}\right]\:, \nonumber\\
& Q_{12} = 4 \, \mathrm{Re}\left[\langle \psi_{0}|\hat{n} \hat{U}^{\dagger}\hat{n} \hat{U}|\psi_{0}\rangle - \langle \psi_{0}|\hat{n}|\psi_{0}\rangle \langle \psi_{0}|\hat{U}^{\dagger}\hat{n} \hat{U}|\psi_{0}\rangle\right]\;,\nonumber\\    
& Q_{21} = 4 \, \mathrm{Re}\left[\langle \psi_{0}|\hat{U}^{\dagger} \hat{n} \hat{U} \hat{n}|\psi_{0}\rangle - \langle \psi_{0}|\hat{n}|\psi_{0}\rangle \langle \psi_{0}|\hat{U}^{\dagger}\hat{n} \hat{U}|\psi_{0}\rangle\right]\;,\nonumber\\
& Q_{22} = 4 \, \mathrm{Re} \left[\langle \psi_{0}|\hat{U}^{\dagger}\hat{n}^{2} \hat{U}|\psi_{0}\rangle - \big(\langle \psi_{0}|\hat{U}^{\dagger}\hat{n} \hat{U}|\psi_{0}\rangle\big)^{2} \right]\;.
\label{qf}
\end{align}

Next, we consider coherent states and squeezed vacuum as the initial probe states \(|\psi_{0}\rangle\),
\begin{align}
    \ket{\alpha} = \hat{D}(\alpha) \ket{0} \quad\mbox{and}\quad 
    \ket{\xi} = \hat{S}(\xi) \ket{0} \,,
\end{align}
where the displacement and squeezing operators are given by 
\(\hat{D}(\alpha) = \exp\left[-\alpha^* \hat{a} + \alpha \hat{a}^{\dagger} \right]\)  and
\(\hat{S}(\xi) = \exp\left[ \frac{1}{2} \left( \xi^* \hat{a}^2 - \xi \hat{a}^{\dagger 2} \right) \right]\),
respectively, and the complex numbers
\( \alpha = |\alpha| e^{i\phi_\alpha} \), and
\( \xi = r e^{i\phi_r} \)
denote the coherent and squeezing amplitudes. The mean energy is given by $\bar n = \sinh^2 r$ and $\bar n = |\alpha|^2$, respectively.

To remove sloppiness, we introduce second and third order scrambling transformations of the form
\begin{equation}
    \hat{U} =e^{- i  \gamma\, \hat{x}^{m}} \qquad (m = 2,3)
\end{equation}
where \(\gamma\) is the scrambling strength, \(\hat{x} = \hat{a} + \hat{a}^{\dagger}\) is the quadrature operator and we take \(m=2\) and \(m=3\) for quadratic and cubic scrambling, respectively. The scrambling strength $\gamma$ governs the degree of parameter mixing. When $\gamma = 0$,  $\hat{U} = \mathbb{\hat{I}}$, and the two phases are perfectly correlated, effectively combining into a single phase parameter $\phi_1 + \phi_2$. This leads to maximal sloppiness, characterized by $\det Q \rightarrow 0$, since only a single function of the phases can be estimated. As $\gamma$ increases, the correlations between $\phi_1$ and $\phi_2$ weaken, allowing $\hat{V}_1$ and $\hat{V}_2$ to encode independent information onto the quantum state. The scrambling operation $\hat{U}$ thus decouples $\phi_1$ and $\phi_2$, enabling control over both the sloppiness of the model and the non-commutativity of their associated SLDs. By investigating this model, we aim to minimize the sloppiness by scrambling the information using nonlinear transformation, possible enhancing precision and reducing quantum incompatibility. We compute the QFIM and the Uhlmann matrix elements for both probe states under both types of scrambling operations, quadratic and cubic. The detailed expressions can be found in Appendix~A, using the explicit form given in Eqs.~(\ref{A}) and (\ref{B}).
\subsection{Addressing Sloppiness}
\begin{figure}[h!]
	\subfloat[\label{fig2a}]{%
		\includegraphics[width=.48\columnwidth]{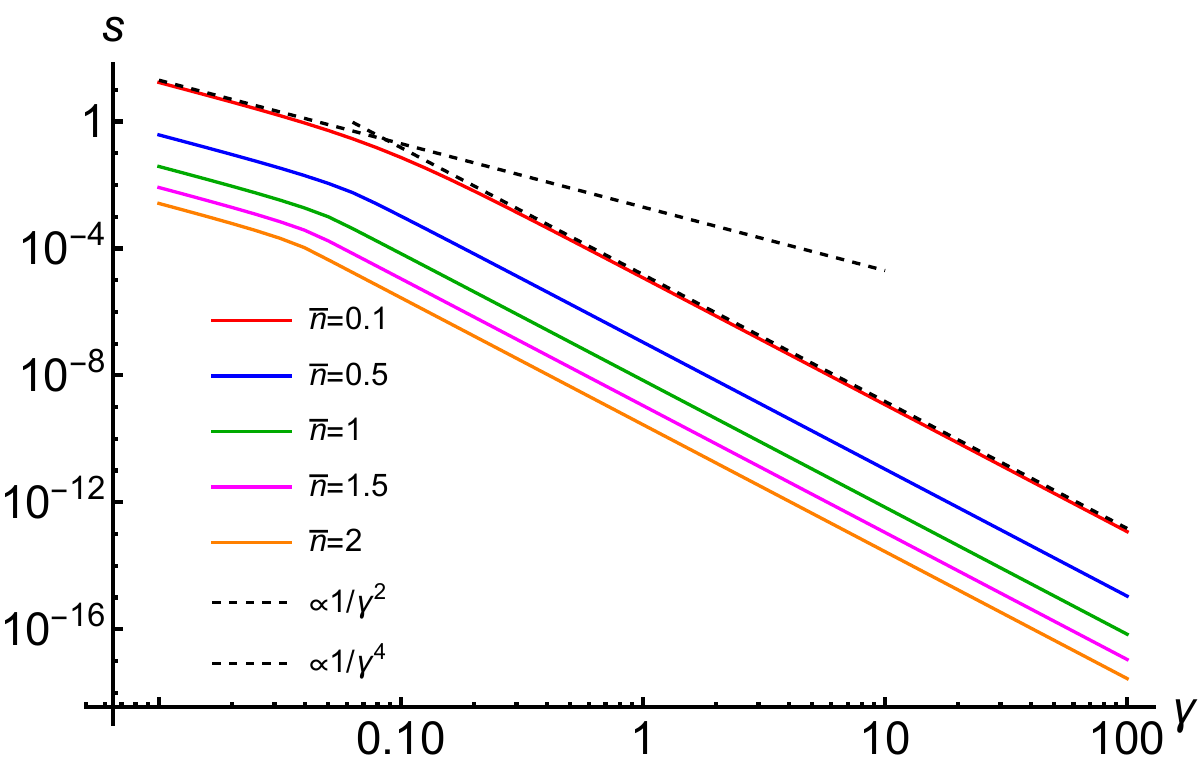}%
	}
	%\hfill
	\subfloat[\label{fig2b}]{%
		\includegraphics[width=.48\columnwidth]{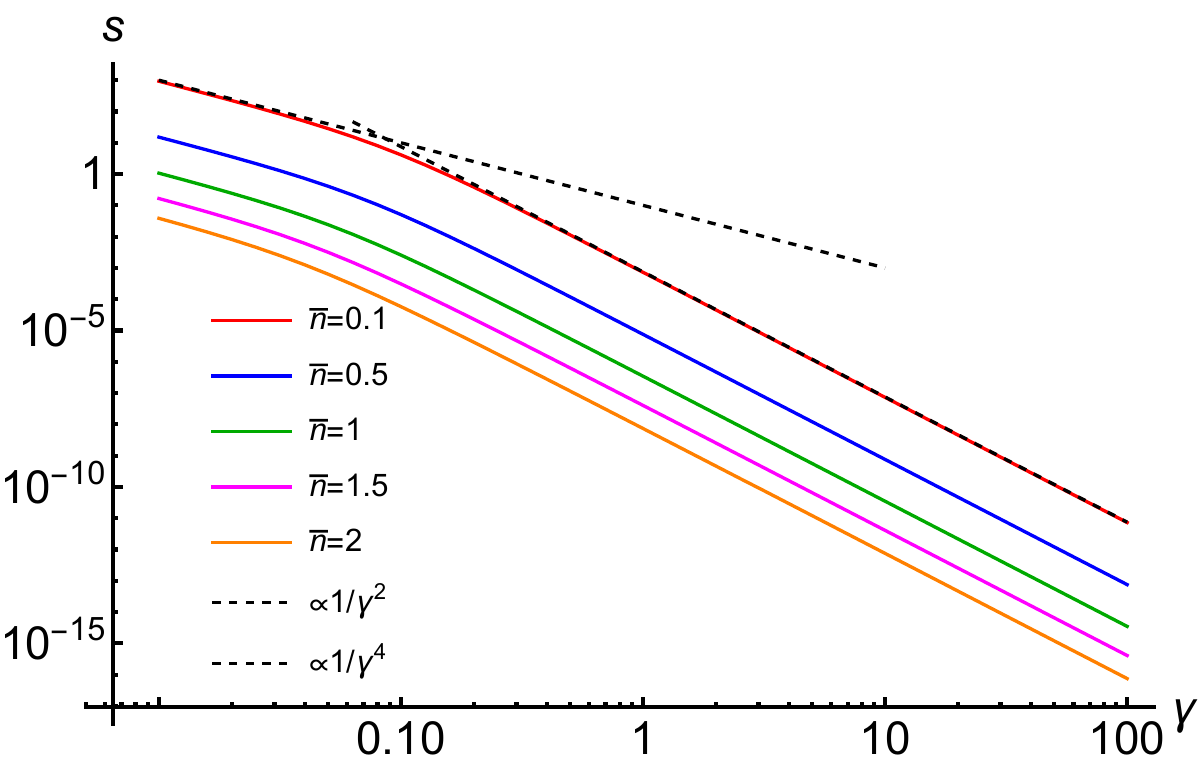}%
	}\\
	%\hfill
	\subfloat[\label{fig2c}]{%
	    \includegraphics[width=.48\columnwidth]{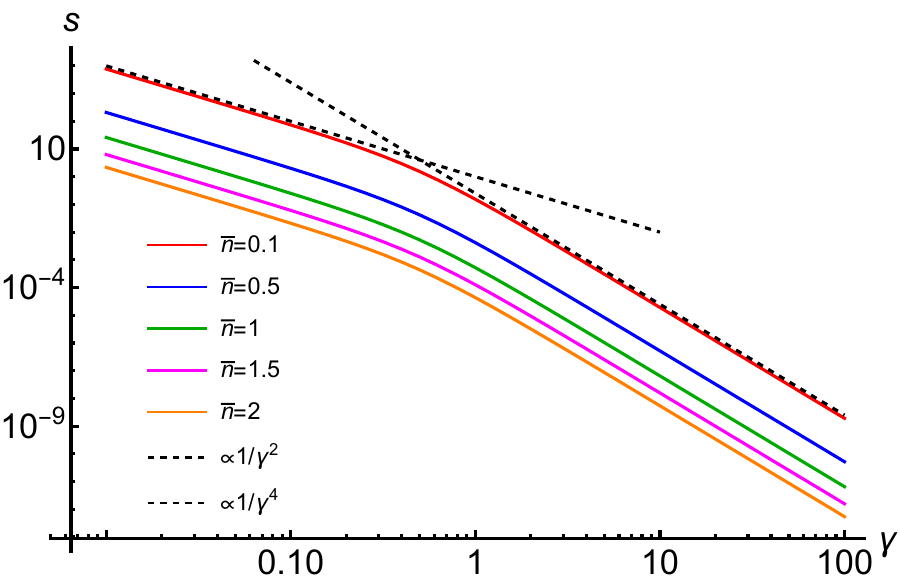}%
        }
        %\hfill
        \subfloat[\label{fig2d}]{%
	    \includegraphics[width=.48\columnwidth]{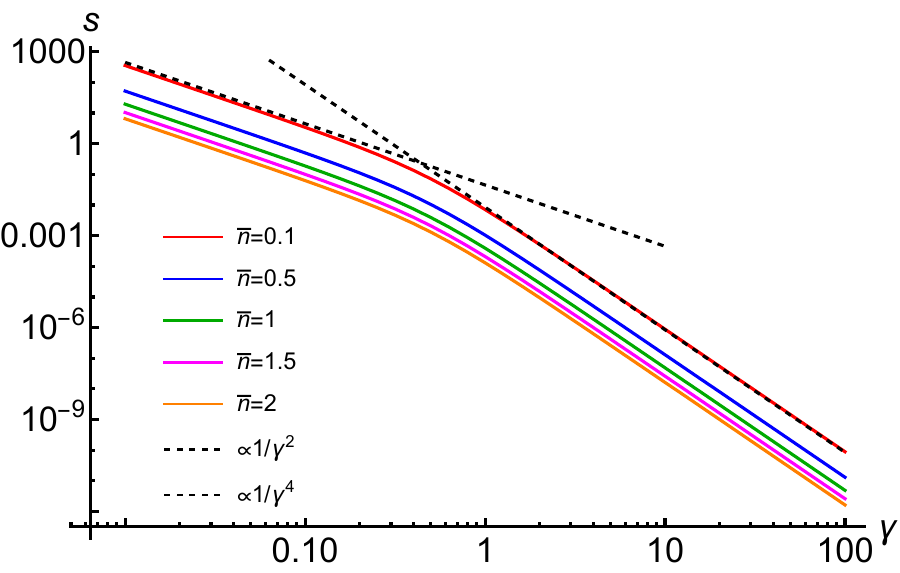}%
        }
\caption{Plots of sloppiness \(S\) versus scrambling strength \(\gamma\): (a) Squeezed vacuum probe with cubic scrambling; (b) Squeezed vacuum probe with quadratic scrambling; (c) Coherent probe with cubic scrambling; (d) Coherent probe with quadratic scrambling.}
\label{S}
\end{figure}

We now address quantitatively how the use of nonlinear scrambling may reduce sloppiness of the model.
In Fig. \ref{S}, we show the results obtained by optimally tuning the phase difference between 
the value of $\phi_1$ and the phase of the signal, either $\phi_\alpha$ or $\phi_r$. 
Figure~\ref{S} shows that the sloppiness \(S\) decreases as a function of nonlinearity \(\gamma\) and
the decreasing is more pronounced for squeezed vacuum than for coherent probe. At fixed \(\gamma\), \(S\) decreases with the energy of the probe (determined by $r$ and $\alpha$, respectively). From Fig.~\ref{S}, we can also see that for fixed value of \(r\) and \(\alpha\), cubic scrambling is more effective than the quadratic one in reducing sloppiness. 
\par
%%%
%A multi-parameter statistical model, classical or quantum, sometimes fails to uniquely associate parameter values with states in a small neighborhood of parameter space. In such cases, different sets of parameters may lead to the same state, making it impossible to distinguish between them from the measurement outcomes: this is known as {\it sloppiness}. Mathematically, sloppiness is indicated by the degeneracy of the FIM in the classical case, or the QFIM, its quantum counterpart. When the matrix is not invertible, it suggests that there is a reparametrization, possibly nonlinear, such that some new parameters contribute very little to the model, while the remaining ones carry most of the relevant information. Formally, if a statistical model can be reparametrized so that it depends on fewer parameters, the QFIM will reflect this through null eigenvalues. Therefore, we define a quantum statistical model to be sloppy if the QFIM is singular, i.e., 
%\(\det Q = 0\). This means that only $m < n$ independent combinations of the original parameters $\phi_1, \phi_2, \dots, \phi_n$ are truly encoded in the quantum states. The eigenvalues of the QFIM indicate the sensitivity of the state to small changes in each parameter direction. On the one hand, a small eigenvalue implies that the quantum state is almost unaffected by changes along that direction, and, thus, that parameter is poorly encoded; on the other hand, a large $S$ indicates strong anisotropy in parameter sensitivity, and the system is sensitive to certain parameter combinations and insensitive to others.

For cubic scrambling, the optimal phase difference is close 
(while not exactly equal) to zero for coherent signal. For squeezed probes, it is close to $\pi/2$ for 
low values of $\gamma$ and to $\pi$ for large values of $\gamma$. For quadratic scrambling, we do not 
have a fixed optimal phase, which rather depends on nonlinear scrambling strength $\gamma$. From the 
analytic expressions of sloppiness for squeezed vacuum and coherent probe,
\begin{subequations}
\begin{align}
 S_{\rm sq}^{(2)} &= \frac{1}{256\, \gamma^2\,\bar n (1+ \bar n)\,  [\cos \phi_{1} + 2\gamma \sin \phi_{1}]^2}\, \label{S2}\\
 S_{\rm ch}^{(2)} &= \frac{1}{128\,  \gamma^2\, \bar n[(1+\bar n)(1+4\gamma^2) + \bar n \left((1-4\gamma^2) \cos 4\phi_{1} + 4\gamma \sin 4\phi_{1}\right)]}\, \label{s2}
 %C_{2 \rm ch} &= \frac{1}{256 \bar n^2 \gamma^{2} (cos 2 \phi_{1} + 2\gamma sin 2 \phi_{1})^2}\ \label{C2}  
\end{align}
\end{subequations}
we see that the optimal phase minimizing the sloppiness is $\phi_{1} - \phi_{\alpha/r}= \arctan (2\gamma)$ for squeezed probes, and $\phi_{1}- \phi_{\alpha/r} = \frac{1}{4} \arctan  \left(\frac{4\gamma}{1-4 \gamma^2}\right)$ for coherent signal. Notice that sloppiness exhibits negligible sensitivity to phase variations within a neighborhood of those values.

For small values of \(\gamma\), we have the following 
expressions of sloppiness for squeezed vacuum and coherent probe, respectively, for the cubic scrambling:
\begin{subequations}
\begin{align}
    S_{\rm sq}^{(3)} &= \frac{1}{288\, f_{\rm sq}(\bar n)\, \gamma^{2}} - g_{\rm sq}(\bar n) + \mathcal{O}(\gamma)  \\
    S_{\rm ch}^{(3)} &= \frac{1}{144\, f_{\rm ch}(\bar n)\, \gamma^{2}}-  g_{\rm ch}(\bar n) + \mathcal{O}(\gamma),  
\end{align}
\end{subequations}
with:
\begin{align}
f_{\rm sq}(\bar n) &=  \bar{n}(1+\bar{n})(1+2\bar{n})[7+60\bar{n}(1+\bar{n})]  \stackrel{{\bar n} \gg 1}{\sim}  {\bar n}^5\,,
\\ 
g_{\rm sq}(\bar n) &= \frac{3(1+\bar{n}+\bar{n}^{2})}{\bar{n}(1+\bar{n})[7+60\bar{n}(1+\bar{n})]^{2}} \stackrel{{\bar n} \gg 1}{\sim}  \frac1{{\bar n}^4}\,,\\ 
f_{\rm ch}(\bar n) &=  \bar{n}[7+8\bar{n}(5+2\bar{n})] \stackrel{{\bar n} \gg 1}{\sim}  {\bar n}^3\,, \\ 
g_{\rm ch}(\bar n) &= \frac{6[1+2\bar{n}(5+6\bar{n})]}{\bar{n}[7+8\bar{n}(5+2\bar{n})]^{2}}
\stackrel{{\bar n} \gg 1}{\sim}  \frac1{{\bar n}^3}\,,
\end{align}
and for the quadratic scrambling:
\begin{subequations}
\begin{align}
    S_{\rm sq}^{(2)} &= \frac{1}{256\, \bar n (1+\bar n)\, \gamma^{2}}-  \frac{1}{64 \bar n (1+  \bar n)} + \mathcal{O}(\gamma)\\
    S_{\rm ch}^{(2)} &= \frac{1}{128\, \bar n (1+ 2 \bar n)\, \gamma^{2}}-  \frac{1}{32 \bar n (1+ 2 \bar n)} + \mathcal{O}(\gamma).    
\end{align}
\end{subequations}
The above expressions capture the fact that when the nonlinearity is small, one can reduce sloppiness
by increasing the energy of the probe, with squeezed vacuum being much more effective, being
$f_{3sq}(\bar n)\sim{\bar n}^5$, than coherent probe, where $f_{3ch}(\bar n)\sim{\bar n}^3$. 
%For quadratic scrambling, we have,
%\begin{align}
 %S_{2\rm sq} &= \frac{1}{f_{2sq}(\bar n)\,\gamma^{2}}-g_{2sq}(\bar n) + \mathcal{O}(\gamma) \nonumber\\
 %   S_{2\rm ch} &= \frac{1}{f_{2ch}(\bar n)\,\gamma^{2}}- g_{2ch}(\bar n)+ \mathcal{O}(\gamma)   
%\end{align}
%where
%\begin{align}
%f_{2sq}(\bar n) &=  128 \bar{n}(1+\bar{n})(1+2\bar{n})[7+60\bar{n}(1+\bar{n})]
%\stackrel{{\bar n} \gg 1}{\sim}  {\bar n}^5\\ 
%g_{2sq}(\bar n) &=  \frac{3(1+\bar{n}+\bar{n}^{2})}{\bar{n}(1+\bar{n})[7+60\bar{n}(1+\bar{n})]^{2}}
%%\stackrel{{\bar n} \gg 1}{\sim}  \frac1{{\bar n}^4}\\ 
%f_{2ch}(\bar n) &=  64\bar{n}[7+8\bar{n}(5+2\bar{n})]
%\stackrel{{\bar n} \gg 1}{\sim}  {\bar n}^3 \\ 
%g_{2ch}(\bar n) &= \frac{6[1+2\bar{n}(5+6\bar{n})]}{\bar{n}[7+8\bar{n}(5+2\bar{n})]^{2}}
%\stackrel{{\bar n} \gg 1}{\sim}  \frac1{{\bar n}^3}\,,
%\end{align}
%leading to the same conclusions we made for the cubic case. 
For large \(\gamma\), we have instead
%\begin{align}
   % s &= \left[ \frac{1}{ 62208 \bar{n}(1 + \bar{n})  + 8 \sqrt{\bar{n}(1 + \bar{n})} + 80 \sqrt{\bar{n}^3(1 + \bar{n})} + 192 \sqrt{\bar{n}^5(1 + \bar{n})} + 128 \sqrt{\bar{n}^7(1 + \bar{n})} + 32 \bar{n} (1 + \bar{n})(1 + 2\bar{n})^2 \gamma^{4} } \right] + \mathcal{O} (\frac{1}{\gamma^{5}})\nonumber\\
    %s &=  \left[ \frac{1}{124416 \bar{n} \left(1 + 2 \bar{n}(5 + 6 \bar{n})\right) \gamma^{4}} \right] + \mathcal{O} (\frac{1}{\gamma^{5}}) , 
%\end{align}
\begin{subequations}
\begin{align}
S_{\rm sq}^{(3)} &= \frac1{
    62208\,\kappa_{\rm sq}^{(3)} ({\bar n})\,\gamma^4 } + \mathcal{O}\left(\frac{1}{\gamma^{5}}\right)\,,\\[1ex]
    S_{\rm ch}^{(3)} &=   \frac{1}{124416 \,\kappa_{\rm ch}^{(3)}(\bar n)\, \gamma^{4}} + \mathcal{O} \left(\frac{1}{\gamma^{5}}\right), 
\end{align}
\end{subequations}
with
\begin{align}
\kappa_{\rm sq}^{(3)} (\bar n) & = \bar{n}(1 + \bar{n}) [1
    + 8 \sqrt{\bar{n}(1 + \bar{n})} + 80 \sqrt{\bar{n}^3(1 + \bar{n})} 
     \nonumber\\
  & \qquad+ 192 \sqrt{\bar{n}^5(1 + \bar{n})}   + 128 \sqrt{\bar{n}^7(1 + \bar{n})} 
    + 32\, \bar{n}(1 + \bar{n})(1 + 2\bar{n})^2]\,,\\
\kappa_{\rm ch}^{(3)}(\bar n)& = \bar{n} \left[1 + 2 \bar{n}(5 + 6 \bar{n})\right]\,.
\end{align}
and
\begin{subequations}
\begin{align}
S_{\rm sq}^{(2)} &= 
\frac1{
   1024\,\kappa_{\rm sq}^{(2)} ({\bar n})\,\gamma^4 }
+ \mathcal{O}\left(\frac{1}{\gamma^{5}}\right) \nonumber\\[1ex]
    S_{\rm ch}^{(2)} &=   \frac{1}{512\,\bar n\, \gamma^{4}}  + \mathcal{O} \left(\frac{1}{\gamma^{5}}\right)\,, 
\end{align}
\end{subequations}
where
\begin{align}
\kappa_{\rm sq}^{(2)}(\bar n)& = \bar{n} (1 +  \bar{n})\,.
\end{align}
The above expressions show that for large $\gamma$ sloppiness decreases with both the nonlinearity and the probe energy. 
%\begin{align}
  %  s &= \left[\frac{1}{ 12288 \bar{n}(1 + \bar{n}) 1 + 8 \sqrt{\bar{n}(1 + \bar{n})} + 80 \sqrt{\bar{n}^3(1 + \bar{n})} + 192 \sqrt{\bar{n}^5(1 + \bar{n})} + 128 \sqrt{\bar{n}^7(1 + \bar{n})} + 32 \bar{n} (1 + \bar{n})(1 + 2\bar{n})^2 \gamma^{4} } \right] + \mathcal{O} (\frac{1}{\gamma^{5}}) \nonumber\\
   % s &= \left[ \frac{1}{24576 \bar{n} \left(1 + 2 \bar{n}(5 + 6 \bar{n})\right) \gamma^{4}} \right] + \mathcal{O} (\frac{1}{\gamma^{5}}), 
%\end{align}
Notice that for small values of $\gamma$ we have $S \propto \gamma^{-2}$, while for large \(\gamma\) the dependence switches to $S \propto \gamma^{-4}$.

\subsection{Bounds to Precision}
%The connection between sloppiness and incompatibility in multiparameter quantum estimation isn’t just theoretical, it can help in improving measurement precision. In this work, we show how to use smart parameter encoding and carefully chosen measurements to optimize precision in a two-parameter model. 
In our model, the two shift parameters are equally important and therefore, we set the weight matrix to 
\(W = I\), the identity matrix. As a consequence, the different quantum 
Cram\'er-Rao bounds in Eq.~(\ref{CH1}) satisfy the relation
%\[s Tr[Q]=C_{Q}\leq C_{H} \leq C_{Q} + 2 s \sqrt{det D} \leq C_{Q}\left(1+\sqrt{\frac{s}{c}}\right) \]
\begin{equation}
 C_{Q}\leq C_{H} \leq C_T=C_{Q} + 2  \frac{S}{\sqrt{C}},
 \label{CH2}
\end{equation}
%\[s Tr[Q]=C_{Q}\leq C_{H} \leq C_{Q} + 2  \frac{s}{\sqrt{c}} \leq C_{Q}\left(1+\sqrt{\frac{s}{c}}\right) \]
From Eq.~(\ref{CH2}), we can see that \(C_{H}\) lies between \(C_{Q}\) and \(C_{T}\) and to minimize \(C_{T}\), we need to jointly minimize sloppiness \(S\) and maximize compatibility, \(C\). In Fig.~\ref{CS}, we show the different bounds, minimized over the values of $\phi_1$ (notice that this phase is in general different from the phase minimizing the sloppiness). From Fig.~\ref{CS}, we conclude that when the nonlinear strength is sufficiently large, the ultimate precision bound for the joint estimation of the two phase shifts coincides with the SLD bound $C_{Q}$. This bound can be achieved using measurements performed on independently prepared probes, regardless of their amplitude. In contrast, for lower values of $\gamma$, a regime that may be relevant in certain scenarios, the gap between $C_{Q}$ and $C_{T}$ becomes more significant. In this case, achieving the ultimate precision bound likely requires collective (entangled) measurements across multiple probe preparations. 

\subsection{Joint vs Step-wise estimation}
So far, we have focused on the simultaneous estimation of the two phases, 
\(\phi_{1}\) and \(\phi_{2}\). We now discuss how joint estimation strategies compares with 
step-wise ones, in which the available copies of the 
probe state are split into two subsets, each one dedicated to estimating only one of the parameters separately \cite{mukhopadhyay2025beating,sharma2025mitigating}. Step-wise strategies involve estimating parameters sequentially, rather than estimating them jointly. Given $M$ repeated preparations of the system, we assume to devote $M/2$ of them to estimate solely $\phi_1$, assuming $\phi_2$ unknown, and the remaining $M/2$ preparations to estimate $\phi_2$ with $\phi_1$ known from the first step. Of course, the roles of the two parameters may be exchanged, and we thus have two strategies of this kind.
\begin{figure}[h!]
	\subfloat[\label{fig3a}]{%
		\includegraphics[width=.48\columnwidth]{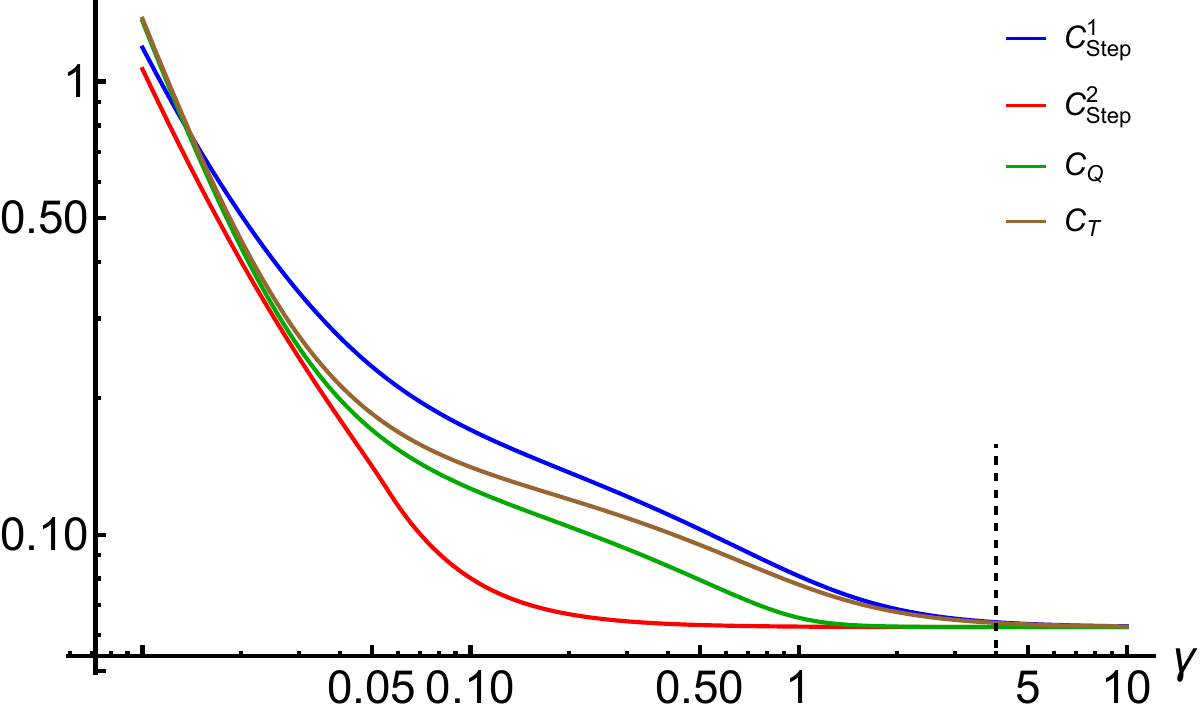}%
	}
	%\hfill
	\subfloat[\label{fig3b}]{%
		\includegraphics[width=.48\columnwidth]{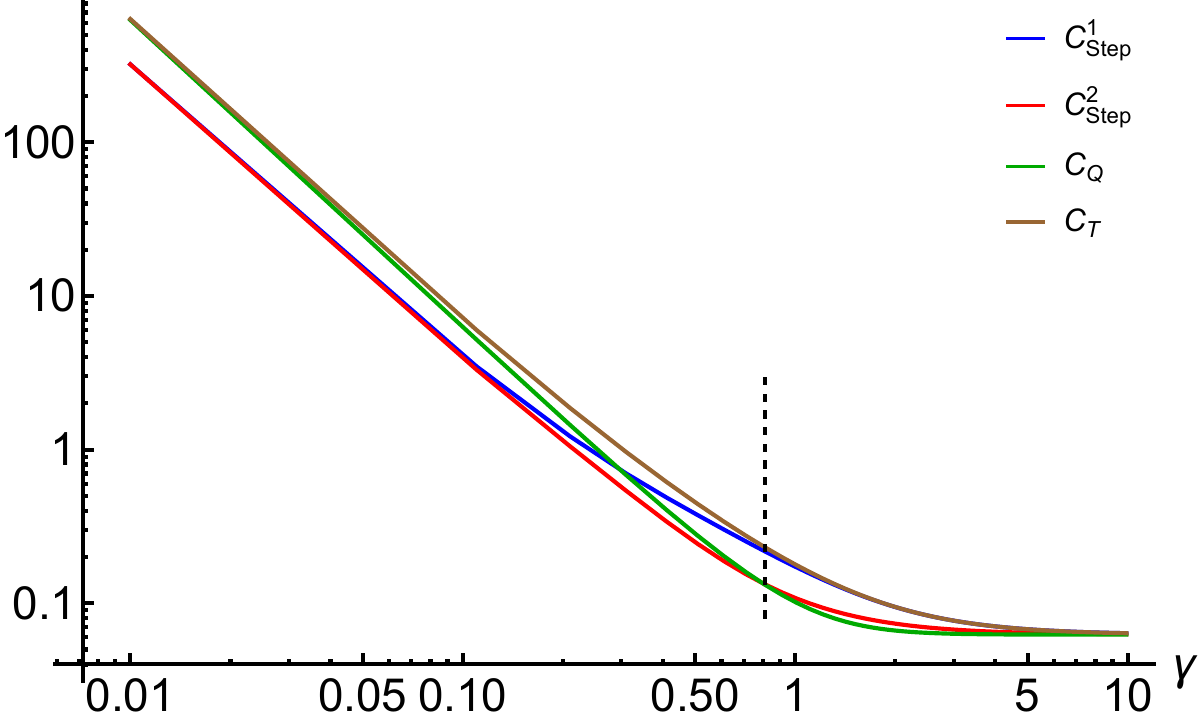}%
	}\\
	%\hfill
	\subfloat[\label{fig3c}]{%
	    \includegraphics[width=.48\columnwidth]{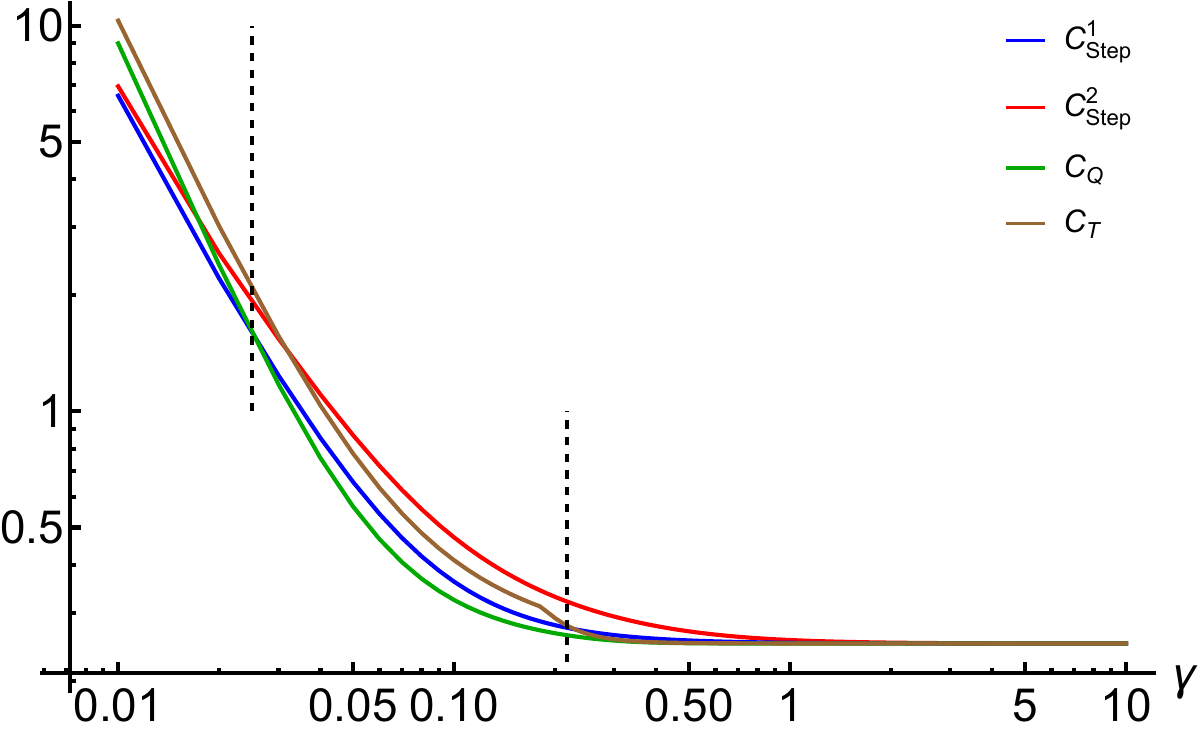}%
        }
        %\hfill
        \subfloat[\label{fig3d}]{%
	    \includegraphics[width=.48\columnwidth]{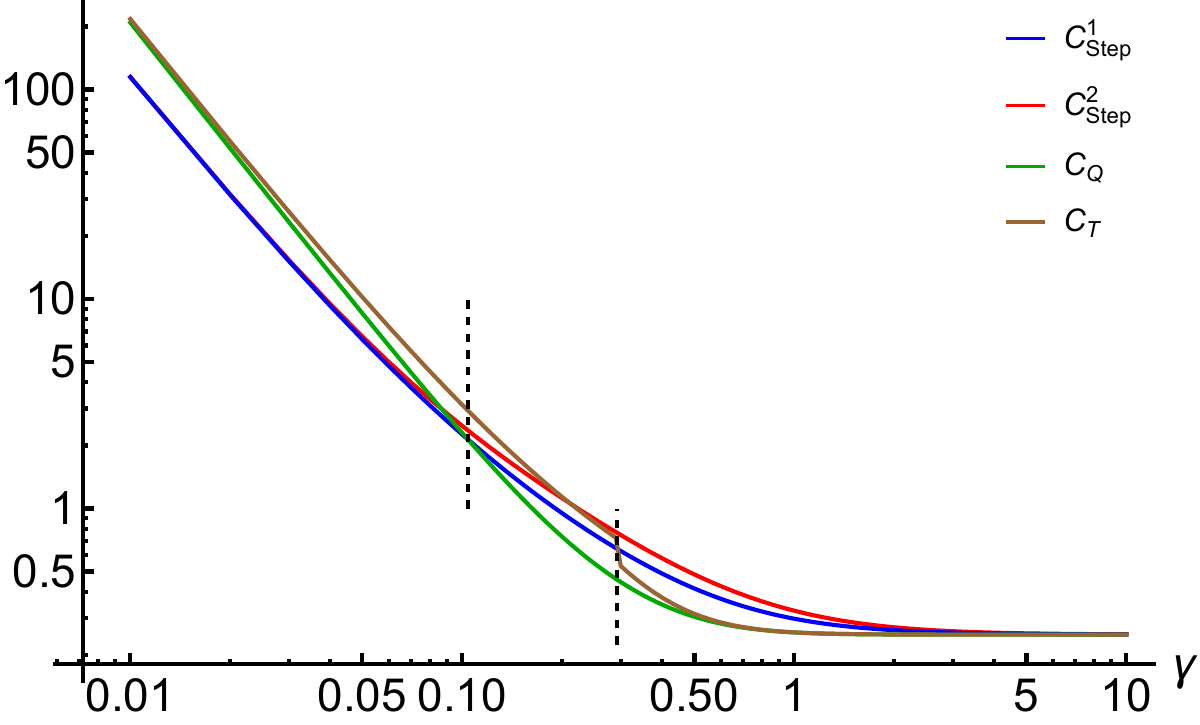}%
        }
\caption{The bounds \(C_{\rm Step}^{1}\), \(C_{\rm Step}^{2}\), \(C_{Q}\) and \(C_{T}\) as functions of the scrambling strength \(\gamma\) for probes with \(\bar{n}=1\): (a) Squeezed vacuum with cubic scrambling; (b) Squeezed vacuum with quadratic scrambling;
(c) Coherent state with cubic scrambling; (d) Coherent state with quadratic scrambling.
For both probes and limited nonlinearity, the best strategy is step-wise estimation,
with cubic scrambling being more effective than quadratic. When squeezed probes and cubic scrambling 
are used, joint estimation becomes superior to step-wise estimation once the nonlinearity passes a certain threshold. For quadratic scrambling, this cannot be proved (see text). For coherent probes, a second threshold on the nonlinearity appears, above which joint estimation outperforms the step-wise, while in the intermediate nonlinearity regime this cannot be proved.}
\label{CS}
\end{figure}
\par
The saturable precision bound on the estimation of $\phi_1$ from the first step is obtained 
from the SLD-QCRB by choosing a weight matrix of the form $W = \mathrm{Diag}(1, 0)$, leading to:
\begin{equation}
    \Delta \phi^{2}_{1} \geq \frac{2\left[Q^{-1}\right]_{11}}{M},
\end{equation}
where $[X]_{ij}$ indicates the elements of matrix $X$.
In the second step, $\phi_1$ is known, and the achievable bound 
to the precision in the estimation of $\phi_2$ is given by the single-parameter QCRB:
\begin{equation}
    \Delta \phi^{2}_{2} \geq \frac{2}{M Q_{22}}
\end{equation}
Therefore, the total variance for this estimation strategy is bounded as:
\begin{equation}
  \Delta \phi^{2}_{1} +\Delta \phi^{2}_{2} \geq \frac{2}{M} \left( [Q^{-1}]_{11} + \frac{1}{Q_{22}} \right)\equiv \frac{C_{\rm Step}^{1}}{M} \,.
\end{equation}
Similarly, reversing the role of the two parameters, we obtain:
\begin{equation}
  \Delta \phi^{2}_{1} +\Delta \phi^{2}_{2} \geq \frac{2}{M} \left( [Q^{-1}]_{22} + \frac{1}{Q_{11}} \right) \equiv \frac{C_{\rm Step}^{2}}{M} \,.
\end{equation}
Notice that, if instead of dividing the total number of repeated preparations equally between the two 
estimation procedures, we had chosen an asymmetric allocation, say $M_1 = \beta M$ measurements 
for estimating $\phi_1$ and $M_2 = (1 - \beta)M$ for estimating $\phi_2$ (or vice versa), with $0\le \beta \le1$, the bounds 
would accordingly be modified.
\begin{equation}
    C_{\rm Step}^{1} = \frac{S Q_{22}}{\beta} + \frac{1}{(1-\beta)Q_{22}}, \quad 
    C_{\rm Step}^{2} = \frac{S Q_{11}}{\beta} + \frac{1}{(1-\beta)Q_{11}}.
\end{equation}
The lowest precision bound among all possible SE schemes is obtained by minimizing 
\(C_{\rm Step}^{1}\) and \(C_{\rm Step}^{2}\) over the possible values of $\beta$, i.e., we have to choose
$
\beta_{\text{opt}} = \arg\min_{\beta} \left( C_{\rm Step}^{1}, C_{\rm Step}^{2} \right)
$. This optimal value is minimizing both \(C_{\rm Step}^{1}\) and \(C_{\rm Step}^{2}\) and is given by
\begin{equation}
    \beta^{k}_{\text{opt}} = \frac{Q_{kk} \sqrt{S}}{Q_{kk} \sqrt{S}+1}\,.
\end{equation}
The corresponding minimized expression are 
\begin{equation}
    C_{\rm Step}^{k,{\rm min}} = \frac{S \left(Q_{kk}+\frac{1}{\sqrt{S}}\right)^{2}}{Q_{kk}},
\end{equation}

In Fig.~\ref{CS}, we show the values of $C_{\mathrm{Step}}^{1}$ and $C_{\mathrm{Step}}^{2}$, alongside $C_Q$ and $C_T$, as functions of the scrambling strength $\gamma$ and for a fixed value of the initial 
energy of the probes ($\bar n=1$). We see that there exist different regimes 
which determines the relative performance of step-wise and joint estimation strategies and cubic scrambling always outperforms quadratic one.  Specifically, step-wise estimation 
is more effective for small values of $\gamma$, whereas for larger nonlinearities, i.e., when scrambling
is enhanced, joint estimation may offer advantages. 

For large \(\gamma\) and squeezed probe, we have
%\begin{align}
\begin{align}
   C_{Q} = C_{T} = C_{\rm Step}^{1} = C_{\rm Step}^{2} & =  \frac{1}{8 \bar{n}(1+\bar{n})} + \mathcal{O}\left(\frac{1}{\gamma^{2}}\right), 
% C_{3Step}^{2} &= \frac{1}{8\bar{n}(1+\bar{n})}+ \mathcal{O}\left(\frac{1}{\gamma^{2}}\right),  \nonumber\\ 
 %C_{2Q} = C_{2T} = C_{2Step}^{1} & =  \frac{(1+2 \bar n)^2}{8 \bar{n}(1+\bar{n})} + \mathcal{O}\left(\frac{1}{\gamma^{2}}\right), \nonumber\\
% C_{2Step}^{2} &= \frac{1}{8\bar{n}(1+\bar{n})}+ \mathcal{O}\left(\frac{1}{\gamma^{2}}\right), 
 \label{SQL}
\end{align}
   %C_{Q}  &= C_{Q} \left(1+T_{Q}\right) = 2\operatorname{csch}^2 2r + \frac{ e^{-4r} (11 - 7 \coth 2r) \operatorname{csch} 2r } {144 \gamma^2} + \mathcal{O}(\frac{1}{\gamma^{3}}), \nonumber\\
 %C_{Step}^{1} &= 2\operatorname{csch}^2 2r - \frac{ e^{4r} (33 - 4\sqrt{3} + 21  \coth 2r) \operatorname{csch} 2r } {432 \gamma^2} + \mathcal{O}(\frac{1}{\gamma^{3}}), \nonumber\\ 
% C_{Step}^{2} &= \frac{\operatorname{csch}^2 2r}{2} + \frac{ e^{4r} \operatorname{csch} 2r \operatorname{sech} 2r } {72 \sqrt{3} \gamma^2} + \mathcal{O}(\frac{1}{\gamma^{3}}), 
%\end{align}
%We can rewrite the above bounds as
%\begin{equation}
   % C_{Q}  = C_{Q} \left(1+T_{Q}\right) = C_{Step}^{1} = \frac{4}{8 \overline{n}(1+\overline{n})} + \mathcal{O}(\frac{1}{\gamma^{2}}), \quad C_{Step}^{2} = \frac{1}{8 \overline{n}(1+\overline{n})} + \mathcal{O}(\frac{1}{\gamma^{2}})
    %\label{SQL}
%\end{equation}
For coherent probes, we have instead
%\begin{align}
% C_{Q}  &= C_{Q} \left(1+T_{Q}\right) = \frac{1}{4 \alpha^{2}} + \frac{1}{3456 \alpha^{2} \gamma^{2}} + \mathcal{O}(\frac{1}{\gamma^{3}}), \nonumber\\
% C_{Step}^{1} &= \frac{1}{4 \alpha^{2}} + \frac{1 + 8 \sqrt{6} }{3456 \alpha \gamma^{2}} + \mathcal{O}(\frac{1}{\gamma^{3}}), \nonumber\\
% C_{Step}^{2} &= \frac{1}{4 \alpha^{2}} + \frac{1 }{72 \sqrt{6} \alpha \gamma^{2}} + \mathcal{O}(\frac{1}{\gamma^{3}}), 
%\end{align}
% Above bounds can be rewritten as
 \begin{align}
     C_{Q} = C_{T}  = C_{\rm Step}^{1} = C_{\rm Step}^{2} &= \frac{1}{4 \overline{n}} + \mathcal{O}\left(\frac{1}{\gamma^{2}}\right), 
     %C_{2Q} = C_{2T} = C_{2Step}^{1} & =  \frac{1}{4 \bar{n}} + 2 + \mathcal{O}\left(\frac{1}{\gamma^{2}}\right), \nonumber\\
     %C_{2Step}^{2} &= \frac{1}{4\bar{n}}+ \mathcal{O}\left(\frac{1}{\gamma^{2}}\right), 
 \label{sql}
\end{align}
Notably, all bounds share the same leading-order term, with differences appearing only in the
subleading terms. From Eqs.~(\ref{SQL}) and (\ref{sql}), we see that the scaling is  shot-noise limited for coherent states \cite{caves1981quantum,PhysRevA.49.3022,giovannetti2004quantum,paris2009quantum} in the large \(\gamma\) limit, whereas squeezing 
allows one to achieve the so-called Heisenberg limit \cite{caves1981quantum,paris1995small}. 

The results illustrated in Fig.~\ref{CS} may be summarized as follows. If one has access only to limited nonlinearity, the best strategy is to employ step-wise estimation, with cubic scrambling being more effective than quadratic. If the available nonlinearity exceeds a certain threshold, joint estimation may becomes more effective: this holds for squeezed probes and cubic scrambling, while for quadratic scrambling cannot be proved since the
best stepwise bound  is between $C_Q$ and $C_T$. In those regime, however, the difference is not dramatic and practical considerations would be the best guideline to choose between stepwise and joint strategies.
\section{Conclusions}\label{s:concl}
In this work, we have analyzed a metrological scheme involving a mode of a bosonic field and 
two successive, unknown phase shifts, which represent the two parameters to be estimated. The corresponding quantum statistical model is intrinsically sloppy, leading to a singular quantum Fisher information matrix (QFIM) and thus making simultaneous estimation of both phase shifts fundamentally challenging. To address this issue, we have introduced an intermediate transformation between the two phase shifts in the form of a nonlinear (quadratic and cubic) scrambling operation. This transformation spreads information across the Hilbert space, thereby mitigating sloppiness and enabling effective multiparameter estimation.

Our analysis confirms that nonlinear scrambling effectively reduces sloppiness, increases parameter compatibility, and enhances the precision of joint estimation. We found that third-order nonlinearity is more effective than second-order. Furthermore, by comparing joint to stepwise estimation, we demonstrated the existence of a threshold for nonlinear coupling, beyond which joint estimation may outperform the stepwise approach if the available nonlinearity is sufficiently large. More specifically, with limited nonlinearity, the optimal strategy is stepwise estimation, where cubic scrambling is more effective than quadratic. If the nonlinearity exceeds a threshold, joint estimation can become more effective, and quadratic scrambling becomes competitive. In the intermediate regime, the performance gap between the two approaches is not pronounced, and practical considerations would be the best guide for choosing between them.

In conclusion, we have proved that nonlinear scrambling is a resource to mitigate or even remove sloppiness of quantum statistical models. By actively manipulating how the information is encoded onto a quantum probe, we can transform an ill-conditioned estimation problem into a viable one. This approach, which trades passive encoding for active control, opens a promising pathway for enhancing quantum multiparameter estimation, where parameter incompatibility and sloppiness are fundamental obstacles.
%%%
\section*{ACKNOWLEDGMENT}
This work received support from MUR and EU through Project  G53D23001110006 {\em Recovering Information in Sloppy Quantum modEls} (RISQUE).
%%%
\section*{Appendix A. Quantum Fisher information matrix and mean Uhlmann curvature}
We report here the explicit expressions of the entries of the QFIM and the incompatibility matrix (mean Uhlmann curvature).
\subsection{Squeezed vacuum probe state and cubic scrambling}
QFIM matrix elements:
\begin{align*}
    Q_{11}(r) &= 8 \cosh^2 r \sinh^2 r \;, \\
    Q_{22}(r, \gamma, \phi_1) &= 
    \frac{1}{2} \Bigg\{
        2\left(-1 + 4374\, \gamma^4 + \cosh4r\right) + 9 \gamma^2 \Big[ 
            5 \cosh2r + 51 \cosh6r + \\
            &\quad\quad 108 \gamma^2 \left(20 \cosh4r + 35 \cosh8r + 128 \cos4\phi_1 \cosh^4 r \sinh^4 r \right) \\
            &\quad\quad - 41 \cos\phi_1 \sinh2r - 3 \cos\phi_1 \left[37 \cosh4r + 576 \gamma^2 \left(9 \cosh2r + 7 \cosh6r\right)\right] \sinh2r \\
            &\quad\quad + 12 \cos2\phi_1 \left[-\cosh2r + 144 \gamma^2 \left(5 + 7 \cosh4r\right)\right] \sinh^2 2r \\
            &\quad\quad + 6 \cos3\phi_1 \left(5 - 1152 \gamma^2 \cosh2r\right) \sinh^3 2r
        \Big]
    \Bigg\}, \\
    Q_{12}(r, \gamma, \phi_1) &= 2 \sinh2r \Bigg\{ \sinh2r + 27 \gamma^2 \Big[ -4 \cos\phi_1 \cosh4r + \left(3 + \cos2\phi_1\right) \sinh4r \Big] \Bigg\}.
\end{align*}

Incompatibility matrix elements:
\begin{equation*}
 D_{12}(r, \gamma, \phi_{1}) = 216 \, \gamma^{2} \sin\phi_{1} \sinh2r \Big[ \cosh2r - \cos\phi_{1} \sinh2r \Big],
\end{equation*}
and $D_{21} = - D_{12}$ and \(    D_{11} = D_{22} = 0 \).
%%%
\subsection{Squeezed vacuum probe state and quadratic scrambling} 
QFIM matrix elements:
\begin{align*}
Q_{11}(r) &= 8 \cosh^2 r \sinh^2 r \;, \\
Q_{22}(r,\gamma, \phi_{1}) &= 4 \Big[
    -\tfrac{1}{4}
    + 2\gamma^2
    + 8\gamma^4
    + \big(\tfrac{1}{4} + 6\gamma^2 + 24\gamma^4\big)\cosh 4r \notag\\
    &\quad
    + 4\gamma^2\big[(-1 + 4\gamma^2)\cos 2\phi_{1} - 4\gamma\sin 2\phi_{1}\big]\sinh^2 2r \notag\\
    &\quad
    - 2\gamma(1 + 8\gamma^2)\big(2\gamma\cos\phi_{1} - \sin\phi_{1}\big)\sinh 4r
\Big],
\\
Q_{12}(r,\gamma, \phi_{1}) &= 2\sinh 2r\Big[
    4\gamma\cosh 2r\big(-2\gamma\cos\phi_{1} + \sin\phi_{1}\big)
    + (1 + 8\gamma^2)\sinh 2r
\Big]\:.
\end{align*}

Incompatibility matrix elements:
\begin{align*}
D_{12}(r, \gamma, \phi_{1}) &= 8 \gamma \Big[\cos \phi_{1} + 2 \gamma \sin \phi_{1}\Big] \sinh 2 r
\end{align*}
and $D_{21} = - D_{12}$ and \(    D_{11} = D_{22} = 0 \).

\subsection{Coherent probe state and cubic scrambling}
QFIM matrix elements:
\begin{align*}
    Q_{11}(\alpha) &= 4 \alpha^{2}, \\
    Q_{12}(\alpha, \gamma, \phi_{1}) &= 4 \bigg\{ \alpha^{2} 
    - 3 \gamma \big[3 \alpha^{3} \sin3\phi_{1} + (3 \alpha^{3} + \alpha) \sin\phi_{1} \big] \\
    &\quad + 9 \gamma^{2} \big[4 \alpha^{4} \cos4 \phi_{1} 
    + (16 \alpha^{4} + 12 \alpha^{2}) \cos2 \phi_{1} 
    + 12 \alpha^{4} + 12 \alpha^{2} \big] \bigg\}, \\
    Q_{22}(\alpha, \gamma, \phi_{1}) &= 4 \bigg\{ \alpha^{2} 
    - 3 \gamma \big[6 \alpha^{3} \sin3\phi_{1} + (6 \alpha^{3} + 2 \alpha) \sin\phi_{1} \big] \\
    &\quad + 9 \gamma^{2} \big[2 \alpha^{4} \cos4 \phi_{1} 
    + (40 \alpha^{4} + 40 \alpha^{2}) \cos2 \phi_{1} 
    + 38 \alpha^{4} + 48 \alpha^{2} + 7 \big] \\
    &\quad - 27 \gamma^{3} \big[32 \alpha^{5} \sin5\phi_{1} 
    + (96 \alpha^{5} + 144 \alpha^{3}) \sin3\phi_{1} 
    + (64 \alpha^{5} + 144 \alpha^{3} + 48\alpha) \sin\phi_{1} \big] \\
    &\quad + 81 \gamma^{4} \big[32 \alpha^{6} \cos6 \phi_{1} 
    + (192 \alpha^{6} + 336 \alpha^{4}) \cos4 \phi_{1} \\
    &\qquad + (480 \alpha^{6} + 1344 \alpha^{4} + 768 \alpha^{2}) \cos2 \phi_{1} 
    + 320 \alpha^{6} + 1008 \alpha^{4} + 768 \alpha^{2} + 96 \big] \bigg\}.
\end{align*}

Incompatibility matrix elements:
\begin{align*}
    & D_{12}\left(\alpha, \gamma, \phi_{1}\right) = -4 \bigg\{ 3 \gamma \left[3 \alpha^{3} \cos{3\phi_{1}}+\left(\alpha^{3} + \alpha\right) \cos{\phi_{1}}\right] + 9 \gamma^{2}\left[4 \alpha^{4} \sin{4 \phi_{1}} + \left(8 \alpha^{4}+ 12 \alpha^{2}\right) \sin{2 \phi_{1}} \right]\bigg\}\;,
\end{align*}
and $D_{21} = - D_{12}$ and \(    D_{11} = D_{22} = 0 \).

\subsection{Coherent probe state and quadratic scrambling} 
QFIM matrix elements:
\begin{align*}
Q_{11}(\alpha) &= 4\alpha^2 \\
Q_{12}(\alpha, \gamma, \phi_{1}) &= 4\left[\alpha^2 - 4\gamma \alpha^2 \sin2\phi_{1} + 8\gamma^2(\alpha^2 \cos2\phi_{1} + \alpha^2)\right] \\
Q_{22}(\alpha, \gamma, \phi_{1}) &= 4\big[ \alpha^2 - 8\gamma \alpha^2 \sin2\phi_{1} + 
8\gamma^2(2\alpha^2 \cos2\phi_{1} + 4\alpha^2 + 1) \\
&\quad - 16\gamma^3 (4\alpha^2 \sin2\phi_{1}) + 16\gamma^4(8\alpha^2 \cos2\phi_{1} + 8\alpha^2 + 2) \big]
\end{align*}

Incompatibility matrix elements:
\begin{align*}
D_{12}(\alpha, \gamma, \phi_{1}) &= -16 \alpha^2 \gamma \left[\cos2\phi_{1} + 2\gamma \sin2\phi_{1}\right]
\end{align*}
and $D_{21} = - D_{12}$ and \(    D_{11} = D_{22} = 0 \).
%\section*{References} 
\bibliography{SPbib}
\end{document}